\documentclass{llncs}

\usepackage{times}
\usepackage{code}

\title{The Design of a COM-Oriented Module System}
\author{Riccardo Pucella}
\institute{Department of Computer Science\\Cornell University\\riccardo@cs.cornell.edu}
\date{}

\newcommand\kw[1]{\textbf{#1}}

\newcommand\ifc[1]{\textit{#1}}
\newcommand\fun[1]{\textit{#1}}
\newcommand\iunknown{\ifc{IUnknown}}

\newcommand{\Cplusplus}{\mbox{C\hspace{-.05em}\raisebox{.4ex}{\tiny\bf++}}}
\newcommand{\COMMENTOUT}[1]{}

\begin{document}
\maketitle

\begin{abstract}
We present in this paper the preliminary design of a module system
based on a notion of components such as they are found in COM. This
module system is inspired from that of Standard ML, and features
first-class instances of components, first-class interfaces, and
interface-polymorphic functions, as well as allowing components to be
both imported from the environment and exported to the
environment using simple mechanisms. The module system automates the
memory management of interfaces and hides the \emph{IUnknown}
interface and \emph{QueryInterface} mechanisms from 
the programmer, favoring instead a higher-level approach to handling
interfaces. 
\end{abstract}

\section{Introduction}

Components are becoming the principal way of organizing software and
distributing libraries on operating systems such as Windows NT. In
fact, components offer a natural improvement over classical
distribution mechanism, in the areas of versioning, licensing and
overall robustness. Many languages are able to use such components
directly and even 
dynamically. On the other hand, relatively few languages are able to
directly create components usable from any language, aside from the
major popular languages such as C, \Cplusplus{} or Java.

Interfacing components in standard programming languages has some
draw\-backs however. Since component models typically do not map
directly to the large-scale programming mechanisms of a language,
there is a paradigm shift between code using external components and
code using internal modular units. Similarly, the creation of
components in standard programming languages is not transparent to the
programmer. Specifically, converting a modular unit of the programming 
language into a component often requires a reorganization of the code, 
especially when the large-scale programming mechanisms are wildly
different from the component model targeted. 

One direction currently pursued to handle the complexity and
paradigm shift of using components in general languages is to
avoid the problem altogether and focus on 
scripting languages
to ``glue'' components together and sometimes even 
create components in a lightweight fashion, by simple
composition. This approach is useful for small tasks and moderately
simple programs, but does not scale well to large 
software projects where the full capabilities of a general language
supporting large-scale programming structures is most useful.

A modern general language for programming in a component-based world
should diminish the paradigm shift required to use components versus using the
language native large-scale programming mechanisms. Moreover, it should be
possible to reason about the code, by having a reasonable
semantic description of the language that includes the interaction with
components.

We explore in this paper the design of a language to address this
issue. We tackle the problem by specifying a language that
uses a notion of components as its sole large-scale programming mechanism,
both external components imported from the environment and internal
components written in the programming language. An internal component
can be exported to the environment as is. The model of components on
which the system is based is the COM model. Our reasons for this were
both pragmatic and theoretical. Pragmatically, COM is widely used and
easily accessible. Theoretically, it is less
object-oriented than say CORBA \cite{OMG93}, and one of our goals is
to explore issues in component-based programming without 
worrying about object-oriented issues. Our proposed module system
subsumes both the \emph{IUnknown} interface and the
\emph{QueryInterface} mechanism through a higher-level mechanism based 
on signature matching. 

We take as our starting point the language Standard ML (SML) \cite{Milner97}. SML is a
formally-defined mostly-functional language. One advantage of working
with SML is that there is a clear stratification between the module
system and the core language. For our purposes, this means that  
we can replace the existing module system with minor rework of the
semantics of the core language. Moreover, the SML module system will
be used as a model in our own proposal for a component-based module
system. Note that this is not simply a matter of implementing COM in
SML, using the abstraction mechanisms of the language. We seek to add specific
module-level capabilities that capture general COM-style abstractions. 

This paper describes work in progress. The work is part
of a general project whose goals are to understand components as a
mean of structuring programs, at the level of our current understanding of module
systems, and to provide appropriate support for components in modern
programming languages.

\section{Preliminaries}

In this section, we review the details necessary to understand the
module system we are proposing. We first describe the COM approach to
component architectures, since our module system is intended to
model it. The description is sketchy, but good introductions
include \cite{Rogerson97,Chappell96} for COM-specific information, and
\cite{Szyperski97} for general component-oriented 
information. We then describe the current module system of SML,
since it provides the inspiration and model for our own module system.

\subsection{Components \`{a} la COM}

COM is Microsoft's component-based technology for code reuse and
library distribution \cite{Rogerson97,Microsoft95}. COM is a binary
specification, and relies on a small number of principles. The
underlying idea of COM is that of an \emph{interface} to an 
object, where an object is just an instance of a component. An
interface is a view of a component. Given a COM object, it 
is possible to query the object to see if it provides the given
interface. If the object indeed provides the interface, it returns a
pointer to the interface, and through this pointer it is possible to
invoke the methods implemented by the interface. Specifically, an
interface is simply a pointer to a table of function pointers (called a
\emph{vtable}), one for each method of the interface.

The identification of components and interfaces is done via globally
unique identifiers: A component is identified by a class
identifier (CLSID), and an interface by an interface identifier
(IID). It is important to note that the CLSID of a component is part
of its formal description. When an application registers a component
with the system (so that other applications can use it), it adds the
CLSID of the component to a system database. Similarly, an interface
identifier is associated formally and permanently with a given
interface. To use a COM component, one need the CLSID of the
component, and the IID of an interface of the component. For example,
the Win32 function \fun{CoCreateInstance} expects a CLSID and an
IID to create an instance of the component with that CLSID, and
returns a pointer to the specified interface (it fails if no such
interface is defined). 

An interface can \emph{inherit} from another interface.  An interface $A$ that inherits from interface $B$ simply
specifies that $B$'s methods appear before the method specified by
$A$ in the vtable of the interface. It really is interface inheritance --- not a word is said about
implementation, which need not be shared by $A$ and
$B$.

A special interface is defined by the COM standard. This interface,
\iunknown{}, is required to be inherited by every other interface. The
interface (simplified for our purposes) is defined as follows in
IDL\footnote{IDL is an interface definition language, a notation used
to describe interfaces. It essentially uses the C notation for types,
augmented with attributes.}:
\begin{centercode}
  interface IUnknown \{
    HRESULT QueryInterface ([in] const IID& iid,
                            [out] void **ppv);
    unsigned long AddRef ();
    unsigned long Release ();
  \}
\end{centercode}
Since \iunknown{} is inherited by every interface, every interface must
defines those functions. They are the heart of the COM technology. The
idea behind \emph{QueryInterface} is that the programmer,
having created an instance of a component and obtained a given
interface $A$, can use the method \fun{QueryInterface} of $A$ to
obtain another interface to the given instance. Various requirements
are made of \fun{QueryInterface}, summarized as follows:
\begin{enumerate}
\item Querying for \iunknown\ form any interface of a given component
always returns the same pointer. 
\item From any interface on a component, it is possible to query for
any other interface provided by the component. 
\end{enumerate}

Point 1 is important because it defines the notion of \emph{object
identity}. The requirement is that no matter which interface to a
given instance one is working with, querying that interface for the
\iunknown{} interface is guaranteed to return a specific pointer,
always the same no matter what interface was used to call
\fun{QueryInterface}. Therefore, if querying for \iunknown{} from two
distinct interface yields the same pointer, one is sure that the two
interfaces are actually to the same instance. Point 2 ensures that
all the interfaces of an instance are 
accessible from any interface of the instance. 

The two final methods in \iunknown{}, \fun{AddRef} and \fun{Release},
are used for memory management of interfaces. COM implements a reference-counting
scheme to manage components. \fun{AddRef} is called whenever a
new pointer to an interface is created, incrementing the reference
count of the interface. \fun{Release} is 
called when a pointer to an interface is not to be used anymore (for
example, before the pointer variable goes out of scope), and simply
decrements the reference count. When the count goes to 0, the memory
associated with the interface can be freed by the system. Although
greatly simplifying memory management, correctly using \fun{AddRef}
and \fun{Release} to prevent memory leaks and dangling pointers to
interfaces is not easy, and the burden of safety is put on the
programmer.

Containment and aggregation are two ways of combining and
reusing components. Containment is straightforward: a component
$C_1$ (outer) is said to \emph{contain} a component $C_2$ (inner) if
$C_1$ uses $C_2$ in its implementation. In other words, $C_1$ is a
client of $C_2$. The only requirement for containment is that upon
initialization, the outer component should initialize the inner
component. Aggregation is specific to COM, and can best be seen as an
optimization of containment. Suppose the outer component $C_1$ wants
to expose an interface actually implemented by the inner component
$C_2$. Using containment, $C_1$ would need to define the interface and 
implement every method call by calling the inner component's
interface. The inefficiency introduced by such indirection is slight,
but if many such interface get redirected, the inefficiency
accumulates. Aggregation is a mean of directly exposing the interfaces
of inner components through to the outer component. An important
property of aggregation concerns object identity: 
the inner component should not be recognizable as a distinct
component. Therefore, both the inner and the outer component must
return the same pointer when a query is made for \iunknown.

\subsection{Modules \`{a} la SML}

\begin{figure*}[t]
\begin{centercode}
\kw{signature} PEANO_SIG = \kw{sig}
  \kw{type} N
  \kw{val} zero : N
  \kw{val} succ : N -> N
\kw{end}

\kw{structure} Peano : PEANO_SIG = \kw{struct}
  \kw{type} N = int
  \kw{val} zero = 0
  \kw{fun} succ (n) = n+1
\kw{end}
\end{centercode}
\caption{Peano arithmetic}
\label{fig:peano}
\end{figure*}

Having presented the COM framework, and delineated the target of our
proposed module system, let us review the basics of the SML module
system \cite{MacQueen84}, our underlying model. It is not necessary
for the reader to have a deep knowledge of SML to understand this
presentation. It is 
sufficient to know that SML is a mostly-functional language, with
first-class functions and a polymorphic type system which is
statically checked: programs that do not type-check at compile-time
are flagged as such and rejected. Excellent introductions to SML are
available \cite{Paulson96,Ullman98,Harper98}. 

The basic elements of the SML module system are structures, signatures 
and functors. A structure is a package of possibly both types and
values into a single unit. A signature is the ``type'' of a
structure. Consider the example
in Figure \ref{fig:peano}, a simple structure defining Peano arithmetic with
its corresponding signature. The structure defines a type $N$ of Peano 
integers, a value for \fun{zero} and a function \fun{succ}. The
structure defines a Peano integer to simply be an integer, and the
zero and successor to be simply $0$ and $+1$. The signature
\fun{PEANO} explicitly specifies the types and values that are
visible outside the structure. A signature \emph{matches} a structure
if the signature consistently denotes a subset of the types and values 
of the structure. Matching a structure with a signature declaring less 
information than the structure is called \emph{signature
ascription}. Suppose one wanted to define a structure like \fun{Peano} 
but that did not have a successor function. One could use signature
ascription to control the visibility, as in
\begin{centercode}
  \kw{structure} Peano2 : \kw{sig} 
    \kw{type} N 
    \kw{val} zero : N
  \kw{end} = Peano
\end{centercode}
This example also illustrate signature matching by inlining a
signature description instead of using a named signature. In SML,
signature matching is by default transparent: although signature
ascription can weed out declarations, it does not hide the
representation of the types. For example, the implementation of
\fun{Peano} uses integers to represent the type \fun{N}. Although the
signature does not specify the representation type, the system will
still accept
\begin{centercode}
  3 + (Peano.zero)
\end{centercode}
as well-typed. In effect, the type \fun{N} is viewed as an
abbreviation for the type integer. In contrast, \emph{opaque matching} 
(using the matching symbol \kw{$:>$}) completely hides whatever
information is not specified in the signature, including
representation types. The above sample would then fail to type-check. 

A functor is a parametrized structure. Suppose one wanted to write a
structure defining elementary algebraic operations on the integers
using Peano arithmetic. Since one may have multiple implementation of
Peano arithmetic, the simplest way would be to parameterize the
structure as follows:
\begin{centercode}
  \kw{functor} AlgOpFun (\kw{structure} P : PEANO_SIG) : \kw{sig}
    ...
  \kw{end} = \kw{struct}
    ...
  \kw{end}
\end{centercode}
which declares the functor \fun{AlgOpFun} to take a structure \fun{P}
matching signature \fun{PEANO\_SIG} as parameter and creating a new
structure using structure \fun{P} in its body. Instantiating a functor 
is simple:
\begin{centercode}
  \kw{structure} AlgOp = AlgOpFun (\kw{structure} P = Peano)
\end{centercode}

\section{Design of the module system}

After reading about the SML module system, one recognizes a strong
similarity between the notion of a structure and the notion of the
instance of a component\footnote{
 Especially if one adheres to the Edinburgh school,
 which advocates the creation of structures exclusively through the 
 application of functors with no arguments.
}. A functor with no argument can be seen as a component,
with the generated 
structure corresponding to an instance, and the notion of containment
and aggregation bear a strong resemblance to functors with
parameters. Of course, this preliminary intuition does not take into 
account interfaces and their behavior under the \fun{QueryInterface}
mechanism. 

In this section, we introduce a module system based on the SML module
system, and providing the notions of components and interfaces. We
impose the following design criteria on our design of the system:
\begin{enumerate}
\item Component instances provide interfaces.
\item Interfaces provide both types and values.
\item Component instances and interfaces are first-class, that is they 
can be passed to and returned from functions, and stored in data
structures.
\item Memory management of interfaces is hidden from the programmer.
\item The \fun{QueryInterface} mechanism is subsumed by syntax.
\item Syntactically and operationally, there is no distinction between 
internal and imported components.
\item Exportable components are easily
characterized and general mechanisms are used to make a component
exportable. 
\end{enumerate}

Criteria 1--2 define the ``architecture'' of the module system, the
relationship between components, interfaces and the core language. Criterion 3 is
required if we want to emulate pointer-based interface
manipulation. Criteria 4--5 are important to ease the use of the
system: memory management under COM, although 
easier than it could be, is still fragile in that the user is
responsible for managing reference counts explicitly (in practice, languages
like \Cplusplus\ encourage the use of smart pointers to alleviate most
of the burden). The \fun{QueryInterface} mechanism is
powerful, but very low-level. We can take advantage of patterns
of use and provide a high-level mechanisms for accessing
interfaces. Finally, criteria 6--7 are mandated by the fact that
the module system will be used as the large-scale programming
mechanism of the language. There should be no difference between code
using an internal component versus an imported component. It is clear
that not every Core SML type is exportable (since the interfaces 
must at the very least be expressible in IDL to be exportable), so restricting the
notion of component to what can be meaningfully exported is too
restrictive for components that we don't want to export, that are only 
used internally.  A simple and elegant way to support exportable
components and unrestricted internally-used components is a must
for a truly usable system. We use signature ascription to achieve
this. 

\subsection{Components and interfaces}

\begin{figure*}[t]
\begin{centercode}
\kw{interface_sig} X_SIG = \{
  \kw{val} fooX : unit -> unit
\}

\kw{interface_sig} Y_SIG = \{
  \kw{val} fooY : unit -> unit
\}

\kw{component_sig} FOO_SIG = \{
  \kw{interface} X : X_SIG
  \kw{interface} Y : Y_SIG
\}

\kw{component} FooComp () : FOO_SIG = \{
  \kw{interface} X = \{
    \kw{fun} fooX () = print "fooX"
  \}
  \kw{interface} Y = \{
    \kw{fun} fooY () = print "fooY"
  \}
\}
\end{centercode}
\caption{Simple component example}
\label{fig:component}
\end{figure*}

Let us give a quick overview of the basic elements of the module system. A
component is defined as providing a set of interfaces. A component has 
a signature, which assigns interface signatures to its interfaces. An
interface defines types and values (including functions). An interface 
signature is simply the ``type'' of an interface. Signature ascription
can be used to thin out interfaces from components or types or values
from interfaces. At the present time, we require signatures to be
named. Component definitions are generative: one
needs to instantiate a component to use it. 

Let us illustrate with a simple example, presented in Figure
\ref{fig:component}. To use component \fun{FooComp}, one first
instantiates it by  
\begin{centercode}
  \kw{val} Foo = FooComp ()
\end{centercode}
and accessing its elements is done using the dot notation, so that
\texttt{Foo.X.fooX ()} prints \texttt{fooX}. Interfaces are
first-class, so it is possible to bind an interface, as in
\begin{centercode}
  \kw{val} FooX = Foo.X
\end{centercode}
which corresponds to accessing the \fun{X}
interface of \fun{Foo}. The type of an interface is simply the name of its
signature, surrounded by $||\cdots||$, so that \fun{Foo.X} has type
\fun{$||$X\_SIG$||$}. Similarly, component instances are first-class,
and their type again is simply the name of their signature, surrounded 
by $|\cdots|$, so that \fun{Foo} has type \fun{$|$FOO\_SIG$|$}. 

As a last remark, we mention that signature matching is opaque. If one 
wants to carry representation types through a signature, one needs to explicitly
give the representation types in the signature, as in
\cite{Harper94,Leroy94}. 

\subsection{Parametrized components}

As the notation for component declarations suggests, every component is
parametrized. In Figure \ref{fig:component}, \fun{FooComp} was a nullary
component, a component with no parameters (hence the \texttt{()} in
the declaration of \fun{FooComp}). Here is a sample parametrized
component:
\begin{centercode}
  \kw{component} BarComp (\kw{val} X : ||X_SIG||
                     \kw{val} Y : ||Y_SIG||) : BAR_SIG = \{
    ...
  \}
\end{centercode}
where \fun{BarComp} is parametrized over interfaces matching
\fun{X\_SIG} and \fun{Y\_SIG}. A simple instantiation would be
\begin{centercode}
  \kw{val} Bar = BarComp (\kw{val} X = Foo.X
                     \kw{val} Y = Foo.Y)
\end{centercode}
passing in the corresponding interfaces from the \fun{Foo} instance of 
\fun{FooComp}.

\subsection{Importing and exporting components}
\label{s:importing}

One key aspect of the COM framework is the possibility of accessing
components written in different languages, and conversely, of
providing components that can be accessed by different languages. Let
us first see how to import a component in our system. We need a
way to define an interface that is imported from the
environment. This is done through an interface signature as in the previous cases, except that
we need to specify the IID of every interface being imported. 
\begin{centercode}
  \kw{interface_sig} IMPORTED_IFC_SIG = \{
    ...
  \} \kw{with_iid} 00000000-0000-0000-0000-000000000000 
\end{centercode}
The requirement being that the signature of an imported component must 
specify an IID for each interface.

Once all the interfaces that are part of the component to be imported
are specified with their IID, we can import the component from the
environment:
\begin{centercode}
  \kw{import} ImportedFooComp : FOO_SIG = \kw{clsid} 
    00000000-0000-0000-0000-000000000000
\end{centercode}
where the component signature \fun{FOO\_SIG} specifies the interface
signatures of the imported component. The component is imported
through its class identifier (CLSID). The 
component so imported can be  instantiated just like a native nullary
component. Note that interface negotiation is done up-front: when a
component is instantiated, it is checked that all the interface
specified in the signature are present.

The converse of importing a component is to export a component. When
exporting a component a program becomes a component server from which
clients can create and instantiate components. Given a component
\cd{BarComp}, one exports it using the declaration:

\begin{centercode}
  \kw{export} BarComp : BAR_SIG \kw{with_clsid} 
    00000000-0000-0000-0000-000000000000
\end{centercode}

The class identifier specified must be a new GUID, as is the rule in
COM programming\footnote{
 We do not specify how such GUIDs are created. Presumably through the
 programming environment. 
}. The component to be exported must be a
nullary component. The component signature must again specify
interface signatures with interface identifiers. 

In order for the exported component to be a valid COM component, its
interface must at least be expressible in IDL. As we are using Core
SML as our core language, we characterize the SML types that can be
naturally represented in IDL via a suitable mapping. One possible
definition follows: we say that a type $\tau$ is
\emph{IDL-expressible} if either of the following holds: $\tau$ is
\fun{int}, \fun{bool} or \fun{real}; $\tau$ is a record type with all field
types IDL-expressible; $\tau$ is an algebraic datatype with the
alternative types all IDL-expressible; $\tau$ is a list with an
IDL-expressible element type; $\tau$ is a component signature; or $\tau$ is
an interface signature. An interface signature $I$ is
\emph{IDL-expressible} if every type it defines is IDL-expressible and
if for every value $v$ of type $\tau$, either of the following holds: $\tau$
is \fun{int},\fun{bool} or \fun{string}; or $\tau$ is a function type of the
form $\tau_1 \rightarrow \tau_2$ with $\tau_1$ and $\tau_2$ either \fun{unit},
IDL-expressible or tuples of IDL-expressible types. 

A key feature of the design is that at export time, one can use
signature ascription to keep only the portions of a component which
are IDL-expressible. The component itself is fully usable from within
the language, while the restricted version is usable from
without. This still requires the programmer to possibly partition the
interfaces into those that are intended to be exported and those that
are not, but at least the underlying framework is the same, and moreover 
the implementation can be shared across the interfaces.

\subsection{Dynamic interface negotiation}

The mechanism of section \ref{s:importing} for importing components
assumes that the interface negotiation is done up-front, when the
component is instantiated. Clearly, this cannot cover all cases of
interest: one may want to use a component that can be either of two
versions, both containing an interface $A$, but one containing an
``optimized'' version of the interface, called $A'$. Clearly, one
should try to use interface $A'$ if it is available,
otherwise downgrade to using $A$. To do this, we introduce a notion of 
dynamic interface negotiation to handle components in a way compatible 
with other languages. 

We provide an interface case construct to dynamically query a
component instance for a given interface: 
\begin{centercode}
  \kw{ifc_case} x 
    \kw{of} FOO => ...
     | BAR => ...
    \kw{else} => ...
\end{centercode}
This form evaluates to the appropriate expression if instance $x$
supports any of \fun{FOO} or \fun{BAR}, or if none of the cases
apply. To fully benefit from this construct, it is useful to introduce 
a primitive operation \kw{instanceOf}. Given any interface $A$,
\kw{instanceOf} $A$ returns the underlying component instance of the
interface.

\section{Discussion}
\label{sec:discussion}

The main question with respect to the preliminary design of our module 
system, as described in the previous section, is the appropriateness
of the model to capture COM-style component-based programming as it
exists. 

First, note that the system subsumes, at least at this point, a module 
system like SML's, modulo an extra level of indirection. A structure
can be seen as an instance with a single interface, and functors are
just components.  Although the design forces us to write all
structures as functor applications, and we need to access the code
indirectly through the interface, one could imagine syntactic sugar that would
give modules an SML flavor. 

Having first-class component instances and interfaces provides most of
the flexibility needed to handle common COM-style programming in a
type-safe manner. For example, first-class interfaces allow the
definition of \emph{interface-polymorphic 
functions}, functions which are polymorphic over a given interface
type. One can for example define a function of type
\begin{centercode}
  \kw{val} foo : ||FOO_INTERFACE|| -> int
\end{centercode}
that can be applied to any interface matching the
\fun{FOO\_INTERFACE} interface signature. Any instance of a
component with such an interface can be used with that function (by
passing in the interface). Since it is always possible to extract the underlying
instance of an interface, one can also return interfaces while keeping
a handle on the underlying instance. One could imagine a more advanced
type system that would record not 
only the type of interface required by such a function, but also the
set of interfaces that are accessed from that interface. We leave
this investigation for the future. We can similarly define
component-polymorphic functions, where one can moreover use the
subtyping relation on components induced by signature ascription to
define say functions to act on any component providing interface
\fun{FOO} and \fun{BAR}. 

Regarding the suitability to interact with the COM framework through
imported and exported component, the
basic notions are by design compatible. The hiding of \iunknown{} and
of \fun{QueryInterface} greatly simplifies both the design and the 
code. Memory management is hidden from the user, using a combination
of automatically generated underlying calls to \fun{AddRef} and
\fun{Release}. and reliance on garbage collection and
finalization, in the style of the Direct-to-COM compiler
\cite{Gruntz97}.

We have not yet carefully investigated the issues of containment and
aggregation. As we mentionned earlier, those composition mechanisms
have a flavor of functor application, but the match is not exact. One
can write a parametrized component, but the parameterization cannot be over
another component (component instances are first-class, but not
components themselves). A component can be parametrized over an instance,
but this then implies that a component has to be instantiated before
being used as a parameter to another component.  One could solve this problem by making components
higher-order, allowing them to be used as arguments to parametrized
components, or returned from such. Higher-order components correspond to higher-order functors in a SML-style module system,
which greatly complicate the module system theory, especially in
regard to the interaction with types \cite{MacQueen94,Harper94,Leroy95}. With
higher-order components, one
could provide syntactic sugar for convenient aggregation and
containment mechanisms. However, when such components are exported,
the issues raised by Sullivan \emph{et al}
\cite{Sullivan97,Sullivan97a} with regard to the rules required to
ensure the legality of COM components arise, and need to be
addressed. This is another area of future work.

A word about implementation is in order. We use two different
implementation mechanisms, one for internal components, one for
imported components, along with specific handling for exported
components. Internal
components are handled in more or less 
the same way modules are handled in current SML implementation, modulo
first-class interfaces and component instances. Imported components 
rely on the underlying \fun{CoCreateInstance} for creation, and
\fun{QueryInterface} for access to interfaces. Instances of imported
components are represented by their 
\iunknown{} interface pointers, allowing for equality
checking. For exported components, the generation of the appropriate layout of the
vtables can be done on the fly, at
export time.

\section{Related work}

A lot of work has been done independently on both module systems 
and component systems, but none quite taking our approach. Most work in
programming language support for components consists of providing access
to components from within a language's abstraction
mechanism. For example, Java \cite{Gosling96} and \Cplusplus{}
\cite{Stroustrup97}  both map components onto 
classes. Similarly, functional approaches pioneered by Haskell
\cite{Peterson97,PeytonJones98,Finne98,Finne99} and then used in OCaml
\cite{Leroy96,Leroy99} and SML \cite{Pucella99} rely on a combination of abstract types
and functions to interface to components. One can write classes
implementing components in Java and \Cplusplus, and using the
functional approach in Haskell, but the notions do not match
exactly. Our approach to studying the problem is to express components 
with a language \fun{explicitely} geared towards talking about components and
interfaces.

The closest language in spirit to our effort is Component Pascal
\cite{Oberon97}, a language providing extensible records that has been 
used to implement COM components. However, the problem 
is again that there is a paradigm mismatch between the structuring mechanisms of the 
language and components. Component Pascal is a modular language for writing
components, but components themselves are not the modularity mechanism 
of the language. 

COMEL \cite{Ibrahim97,Ibrahim98} was our original inspiration for this 
work. The mechanism of hiding \iunknown{} and \fun{QueryInterface} are 
similar, but the difference is that COMEL is a small language meant for 
a study of the formal properties of the COM framework, while our
proposal is for an extension of an existing language to support
component-based code structuring mechanisms.

\section{Conclusion}

We have presented in this paper a preliminary design for a module
system that directly incorporates the notions of components and
interfaces as defined in COM. The design is rough, but the basic idea
is clear. The system can subsume a powerful module system such as  
SML's, and is directly compatible with COM's model of the world. Work
is now needed to complete and evaluate the design. Aside from the  
issues raised in the text concerning aggregation and higher-order
components, we are working on a formal semantics for the module
system, both a static semantics (types and their propagation) and
a dynamic semantics (the execution model and interaction with the COM
runtime system). The implementation has to be completed and systems built
using it. 

Finally, and this is once again future work, it is of great 
interest to investigate the relationship between our approach and the
approach supported by structuring mechanisms such as units
\cite{Flatt99} and mixins \`a la Jigsaw \cite{Bracha92}.

\textbf{Acknowledgments}. Thanks to Greg Morrisett for many
discussions on module systems and COM. Thanks to Peter Fr\"ohlich for
a thorough reading of the paper and many comments that helped shape
the presentation. Thanks to Dino Oliva, Erik Meijer and Daan
Leijen for early discussions about various aspects of COM. Thanks as 
well to John Reppy for general inspiration.

\appendix

\section{Microsoft Agent}
\label{sec:agent}

A popular example in the literature concerned with interacting with
COM is Microsoft Agent \cite{Microsoft97}. It is an interesting
example because it is simple yet non-trivial, and allows for nice
demonstrations. Microsoft Agent provides a server in charge of 
loading little animated characters that can interact on the
screen. The Agent Server is a component with a single 
interface, \fun{IAgent}. Here is first the Agent Server component
signature, and import from the environment. 

\begin{centercode}
  \kw{component_sig} AGENT_SERVER = \{
    \kw{interface} IAgent : I_AGENT
  \}

  \kw{import} AgentServer () : AGENT_SERVER = \kw{clsid} 
    A7B93C92-7B81-11D0-AC5F-00C04FD97575
\end{centercode}

Here is an extract of the \fun{I\_AGENT} interface signature. Note
that the method \fun{getCharacter} returns an \fun{AGENT\_CHARACTER}
which is as we shall see a component instance that represent a
character that can be animated.  

\begin{centercode}
  \kw{interface_sig} I_AGENT = \{
    \kw{val} load : string -> (int,int)
    \kw{val} unload : int -> unit
    \kw{val} register : |AGENT_NOTIFY_SINK| -> sinkID
    \kw{val} unregister : sinkID -> unit
    \kw{val} getCharacter : int -> |AGENT_CHARACTER|
    ...
  \} \kw{with_iid} A7B93C91-7B81-11D0-AC5F-00C04FD97575 
\end{centercode}

A character is implemented by its own component, with the following
signature. We concentrate on the \fun{IAgentCharacter}
interface. Other interfaces are available to intercept and process
commands to the characters, but we will not be considering those in
this example. 

\begin{centercode}
  \kw{component_sig} AGENT_CHARACTER = \{
    \kw{interface} IAgentCharacter : I_AGENT_CHARACTER
  \}
\end{centercode}

We do not need to import the corresponding component from the
environment, since the creation of characters is restricted to the
\fun{getCharacter} function of the Agent Server component. Indeed, the
\fun{AgentCharacter} component does not explicitly exist in Microsoft Agent.

The \fun{IAgentCharacter} interface is used to control a character, to
make it appear, move about the screen, speak and animate.

\begin{centercode}
  \kw{interface_sig} I_AGENT_CHARACTER = \{
    \kw{val} setPosition : int * int -> unit
    \kw{val} getPosition : unit -> (int * int)
    \kw{val} play : string -> int
    \kw{val} stop : int -> unit
    \kw{val} show : bool -> int
    \kw{val} speak : string * string -> int
    ....
  \} \kw{with_iid} A7B93C8F-7B81-11D0-AC5F-00C04FD97575 
\end{centercode}

The simplest example of code using such an interface is the following, 
which simply displays an agent for 10 seconds.

\begin{centercode}
  \kw{fun} test () = \kw{let}
    \kw{val} AS = AgentServer ()
    \kw{val} (charId,_) = AS.IAgent.load ("merlin")
    \kw{val} Char = AS.IAgent.getCharacter (charId)
  \kw{in}
    Char.IAgentCharacter.show (0);
    Char.IAgentCharacter.speak ("Hello world","");
    sleep (10000);  (* wait for 10000 milliseconds *)
    AS.IAgent.unload (charId)
  \kw{end}
\end{centercode}

We leave the task of defining abstraction and combinators to help
dealing with characters in a sane way, in the style of
\cite{PeytonJones98}, as an exercise to the reader.

\end{document}